\documentclass[reprint,secnumarabic,amssymb, nobibnotes, aps, amsmath, amssymb, prl,groupedaddress,superscriptaddress,longbibliography]{revtex4-1}

\usepackage{graphicx}
\usepackage{dcolumn}
\usepackage{bm}
\usepackage{hyperref}
\usepackage[mathlines]{lineno}
\usepackage{inputenc}
\usepackage{float}
\usepackage{siunitx}

\usepackage[toc,page]{appendix}

\begin{document}

\preprint{APS/123-QED}

\title{Proton acceleration with intense twisted laser light} 

\author{Camilla Willim}
\author{Jorge Vieira}%
\affiliation{%
  GoLP/Instituto de Plasmas e Fus\~ao Nuclear, Instituto Superior T\'ecnico, Universidade de Lisboa, 1049-001 Lisbon, Portugal
}%


\author{Victor Malka}
\affiliation{Department of Physics of Complex Systems, Weizmann Institute of Science, Rehovot 7610001, Israel}%

 \author{Lu\'is O. Silva}
\affiliation{%
  GoLP/Instituto de Plasmas e Fus\~ao Nuclear, Instituto Superior T\'ecnico, Universidade de Lisboa, 1049-001 Lisbon, Portugal
}


\date{\today}

\begin{abstract}
An efficient approach that considers a high-intensity twisted laser of moderate energy (few J) is proposed to generate collimated proton bunches with multi-10-MeV energies from a double-layer hydrogen target. Three-dimensional particle-in-cell simulations demonstrate the formation of a highly collimated and energetic ($\sim 40$ MeV) proton bunch, whose divergence is $\sim 6.5$ times smaller compared to the proton bunch driven by a Gaussian laser containing the same energy. Supported by theoretical modeling of relativistic self-focusing in near-critical plasma, we establish a regime that allows for consistent acceleration of high-energetic proton bunches with low divergence under experimentally feasible conditions for twisted drivers.
\end{abstract}

\maketitle


\section{Introduction}
High-power lasers to generate high-quality ion beams may lead to compact and relatively low-cost sources. Motivated by this argument and by the curiosity to investigate new physical phenomena, various laser-driven acceleration mechanisms have been proposed \cite{macchi,schreiber} such as target normal sheath acceleration (TNSA) \cite{hatchett, snavely,pukhov1}, radiation pressure dominant acceleration (RPA) \cite{rpa_rob,rpa_henig}, collisionless shock acceleration \cite{shock}, and relativistic induced transparency (RIT) \cite{RIT_henig,hybridRPA,relattransp}. Laser-driven ion sources have a wide range of applications \cite{roth_ign, fuchs,patel_heating,bulanov_med,malka_med}. Many of these applications require energies ranging from 1-300 MeV, $\sim$ ps bunch duration, and 1-10 degrees divergence with $10^{11}-10^{13}$ ions in a single bunch. In some cases, plasma-accelerated protons are already in use \cite{jahn2018, neutron_horny}. Still, other impactful applications such as `fast ignition' for inertial fusion or hadron therapy require improved proton sources that deliver strongly collimated proton bunches with energies of tens of MeV to hundreds of MeV \cite{zeil2013,karsch2017,passoni2019}. 
Upgraded schemes to address ion beam properties such as the energy \cite{sgattoni2012,naka_magvortex,wang2013,enh_Zou,arefiev2016,slowlight,DLT_Bin,wan2019, pazzaglia2020, madouble,passoni2016} or the divergence \cite{willingale2006,bin2013,bartal2012,weichman2020} have demonstrated encouraging results. 
Nevertheless, enhancing the ion beam energies while maintaining a few degrees of beam divergence is still an inherent issue of most acceleration mechanisms with laser energies that are, at most, a few tens of Joule. 

The most conventional approach to laser-based ion acceleration employs Gaussian beams. At the same time, there are other types of beams, namely Laguerre-Gaussian beams, which carry a well-defined orbital angular momentum (OAM) along the propagation axis \cite{allen}, that have the potential to enhance the ion beam quality. A Laguerre-Gaussian mode is defined by a radial index $p$ and an azimuthal index $\ell$, which is related to the vortex structure of the beam and quantifies its topological charge. When $\ell \neq 0$, the laser beam consists of a hollow intensity distribution and helical wavefronts, often called twisted light.
Such twisted lasers open new research in highly non-linear ($I\gtrsim 10^{22} \, \text{W/cm}^2$) laser plasma interactions \cite{pariente2015,leblanc2017} including magnetic field generation \cite{shi2018}, electron/positron acceleration in laser-plasma accelerators \cite{vieiraprl2014, vieiraprl2016}, and direct laser acceleration of ions \cite{saberi2017}. 

Using OAM laser pulses to improve laser-plasma ion accelerators has also been proposed using solid targets (single-layer). A divergence reduction was experimentally demonstrated to some extent ($\approx13\%$) through TNSA \cite{brabetz2015}. Furthermore, orbital angular momentum transfer to protons and their compression has been shown through simulations \cite{wang2015}. Recent simulations considering ultra-high laser intensities (I$\gtrsim \, 10^{22}-10^{23}$ W/cm$^{-2}$) have shown proton acceleration of up to GeV energies in a collimated manner, either by accelerating a witness bunch \cite{zhangpro2014} or through the RPA and RIT mechanisms to proton bunches with a divergence angle of $17 \mathrm{mrad}$  ($\sim 10^7$ protons) \cite{pae2020}, and $70 \mathrm{mrad}$ ($\sim 10^{10}$ protons) \cite{ju2021}.
For a target to be opaque, the relativistic transparency factor is $\overline{n} > 1$, where $\overline{n} = n_{e}/\gamma n_{cr}$ and depends on the relativistic Lorentz factor $\gamma = \sqrt{1+a_0^2/2}$ (with normalized laser vector potential $a_0 = e A/m_e c$), the electron density $n_e$ and the critical plasma density $n_{cr}= m_e \epsilon_0 \omega^2 / e^2$. Here, $m_e$ is the electron rest mass, $\epsilon_0$ is the vacuum permittivity, $\omega$ is the angular laser frequency, and $e$ is the electron charge. Double-layer targets made of a thin solid foil preceded by a near-critical plasma ($\overline{n} < 1$) layer are feasible target designs to overcome limitations such as a low conversion efficiency from laser energy into energetic ions at moderate laser intensities.
Theory and experiments demonstrated a remarkable enhancement in ion beam energies using Gaussian laser pulses \cite{sgattoni2012,wang2013,enh_Zou,DLT_Bin,wan2019, pazzaglia2020,madouble,passoni2016} but have not reported a significant decrease in beam divergence. 
 The configuration remains entirely unexplored, using twisted light as a driver. 

In this work, we propose the approach of a double-layer target with twisted light to generate collimated multi-10-MeV proton bunches using moderate laser energies (with I$\simeq \, 10^{20}$ W/cm$^{-2}$). We exploit the effective reduction in proton beam divergence due to a driver with OAM and the enhancement of proton energy due to a double-layer target configuration. We will demonstrate through three-dimensional particle-in-cell simulations and analytic considerations that the laser dynamics in the near-critical plasma part - most notably relativistic self-focusing - play a crucial role in our findings and depend strongly on its OAM contents. Relativistic self-focusing is one of the essential features of the laser interaction with a double-layer target and occurs when ultra-high intensity lasers propagate through a dense plasma. The optimum length of the near-critical plasma layer for enhanced proton energies is the laser's self-focusing length \cite{enh_Zou, pazzaglia2020} due to an efficient hot electron generation which we observe to be similar for both laser drivers. We present an analytical description of the difference in spot size evolution of Gaussian and OAM drivers and show that it can lead to contrasting target responses; and, therefore, proton acceleration mechanisms.
These laser dynamics are crucial for the generation of collimated proton bunches.
We will show that our proposed setup with an OAM driver reduces the divergence of protons up to a factor of $\sim 6.5$ while preserving energy gain and charge ($40 \pm 10\%$ MeV with $0.3$ nC for laser pulses with $3.7$ J) in comparison to a Gaussian driver. Lastly, we identify for a broad range of laser pulse energies ($2-33$ J), a simplified relation to the target composition, i.e. length and densities of the layers, that allows for consistent generation of highly energetic proton bunches (up to 200 MeV) with $>10^{10}$ protons within an ultra-low divergence angle ($\simeq 0.7$ mrad), driven by twisted light.

\section{Relativistic self-focusing in the near-critical plasma layer}
To examine relativistic self-focusing in the near-critical plasma section, we consider the equation for the laser spot size in the highly relativistic limit, $a_0 \gtrsim 1$. The equation of the laser beam envelope may be written as \cite{esarey1997,hafizi2000}
\begin{equation}
\label{eqn:dgleq}
\frac{d^2 X}{dz^2} + (\hat{a}^2 Z_R)^{-2} \frac{\partial V}{\partial X} = 0,
\end{equation}
where the effective potential $V$, defined as 
\begin{equation}
\label{eqn:enveq}
\begin{split}
\frac{\partial V}{d X} = &-16 \frac{P}{P_c} X\Big\{1-(1+X^{-2})^{1/2}-2 \ln{2} + \\
& 2 \ln\left[ 1 + (1+X^{-2})^{1/2}\right]\Big\} - 
 \frac{\ln{\left(1+X^{-2}\right)}}{X} - \frac{1}{X^3},\\
\end{split}
\end{equation}
includes the effects of relativistic focusing, ponderomotive channeling and vacuum diffraction. 
The scaled spot size is defined by $X = w/(a_0 w_0)$, with $a_0$ and $w_0$ as the vacuum amplitude and the spot size at focus, respectively, and $Z_R = k w_0^2/2$ as the Rayleigh length in a vacuum.
The envelope equations for the laser spot size Eqs. (\ref{eqn:dgleq})-(\ref{eqn:enveq}) describe the axial evolution of the spot size as a function of the ratio of the laser power $P$ to the critical power for relativistic self-focusing $P_c$. The critical power ratio takes into account the nature of the mode and is defined as $P/P_G = (k_p a_0 w_0)^2/32$ for a linearly polarized Gaussian beam \cite{esarey1997}, where $k_p$ is the plasma wave number. For the particular case of a single ring structure, where the radial index $p=0$ and the azimuthal index $\ell \neq 0$, the critical power ratio scales with the topological charge $\ell$ of the twisted laser in the following way $P/P_c = \frac{(2\vert \ell \vert) !}{\vert \ell \vert ! (\vert \ell \vert +1) !} \frac{1}{4^{\vert \ell \vert}} P/P_G$ \cite{kruglov1992, jorgeselffocus}. This work focuses on an azimuthal index $\ell = 1$ such that the critical power ratio is $P/P_c = \frac{1}{4} P/P_G$ and therefore, the ratio reduces by a factor of four in comparison to a Gaussian beam. Hence, the OAM light will undergo a significantly less pronounced self-focusing than a pure Gaussian mode which is the crucial property that enables reducing the accelerated proton bunch divergence. During relativistic self-focusing of a Gaussian laser,  the laser undergoes a  corresponding enhancement of its intensity, which can be followed by "ponderomotive" blowout, leading to channeling and filamentation at the edge of the channel walls \cite{cattani2001,mori1988}. When the Gaussian laser amplitude is close to fulfilling the relativistic transparency condition of the thin over-critical plasma layer of the target ($\overline{n}\lesssim 1$), the proton acceleration is influenced by the RPA mechanism with the onset of RIT \cite{hybridRPA,relattransp}, potentially leading to higher proton bunch energies but also a broader divergence. 
 However, the OAM laser amplitude can stay small enough for the overdense plasma layer to stay opaque ($\overline{n}>1$), and so leaves a cold rear surface which benefits lower proton bunch divergence.

We can choose the length, density, and position of the near-critical layer such that an OAM beam consistently drives low divergence proton bunches similar to the TNSA mechanism while the Gaussian beam, containing the same energy, produces broad divergence proton bunches through a hybrid TNSA-RPA mechanism; assuming a thin solid part of the double-layer target and a relativistic transparency factor of $\overline{n} \approx \left\lbrace 1-3\right\rbrace$. A thin solid layer with a thickness on the order of the laser wavelength $\lambda_0$ allows for a better conversion efficiency from laser to protons. We solved the envelope equations Eqs. (\ref{eqn:dgleq}) and (\ref{eqn:enveq}) numerically for different values of $a_0 > 1.0$ and $n_e < n_{cr}$ and found that the length of the near-critical plasma layer $L_1$ should be on the order of the self-focusing length $z_f$, i.e. $L_1 \simeq z_f$. For longer distances, the OAM beam diverges again, and so does the collimated hot electron bunch that moves centered with the driver, in agreement with other studies on double-layer targets  \cite{pazzaglia2020,wang2013}.
We found a focal spot position in the center of the near-critical plasma layer to be favorable.
 A comparison of the dependencies of the relativistic self-focusing behavior on the plasma density, laser amplitude, and position of the focal spot is presented in the appendix. 

\section{Three-dimensional simulations of double-layer target with twisted and Gaussian laser drivers}
We designed a set of three-dimensional PIC simulations (using the PIC code OSIRIS \cite{osiris}) with experimentally feasible parameters that reveal the strong differences between the non-linear propagation of Gaussian and OAM pulses, which confirm our predictions above, based on numerical solutions of Eqs. (\ref{eqn:dgleq}) and (\ref{eqn:enveq}). The double-layer target is composed of hydrogen plasma with a near-critical layer with density $n_1 = 0.4 \, n_{cr}$ and length $\sim 28.6 \, \mu m$ (close to the self-focusing length) preceding a thin overdense plasma layer with density $n_2 = 16 \, n_{cr}$ and length $\sim 1 \, \mu m$. The linearly polarized laser pulses contain the same energy of $E_L \approx 3.7$ J with a duration of $\tau \approx 45$ fs and wavelength $\lambda_0 = 800$ nm. The peak value of the dimensionless amplitude $a_0 = 8.5 \times 10^{-10} \lambda [\mu m] I^{1/2} [\text{W/cm}^2]$ of the laser pulses is set to $a_0 \approx 12$, where $a_0$ is the normalized laser amplitude. To keep the same weighted averaged amplitude $\int_0^\infty \vert a(r) \vert^2 \cdot r dr/\int_0^\infty \vert a(r) \vert^2 dr$ we set the beam waist at $2.4 \, \mu m$ in the OAM case and $4.8 \, \mu m$ in the Gaussian case. The focal spot lies in the center of the near-critical plasma layer. The simulation spatial resolution is $\Delta x = \Delta y = \Delta z \approx \lambda_0 / 31$ with $2$ particles per cell. Due to strong similarities in the $y$ and $z$ directions of the transverse plane, our analysis focuses only on the laser polarization direction $y$ and the longitudinal direction $x$.

\begin{figure}[h!]
\includegraphics[scale = 0.285]{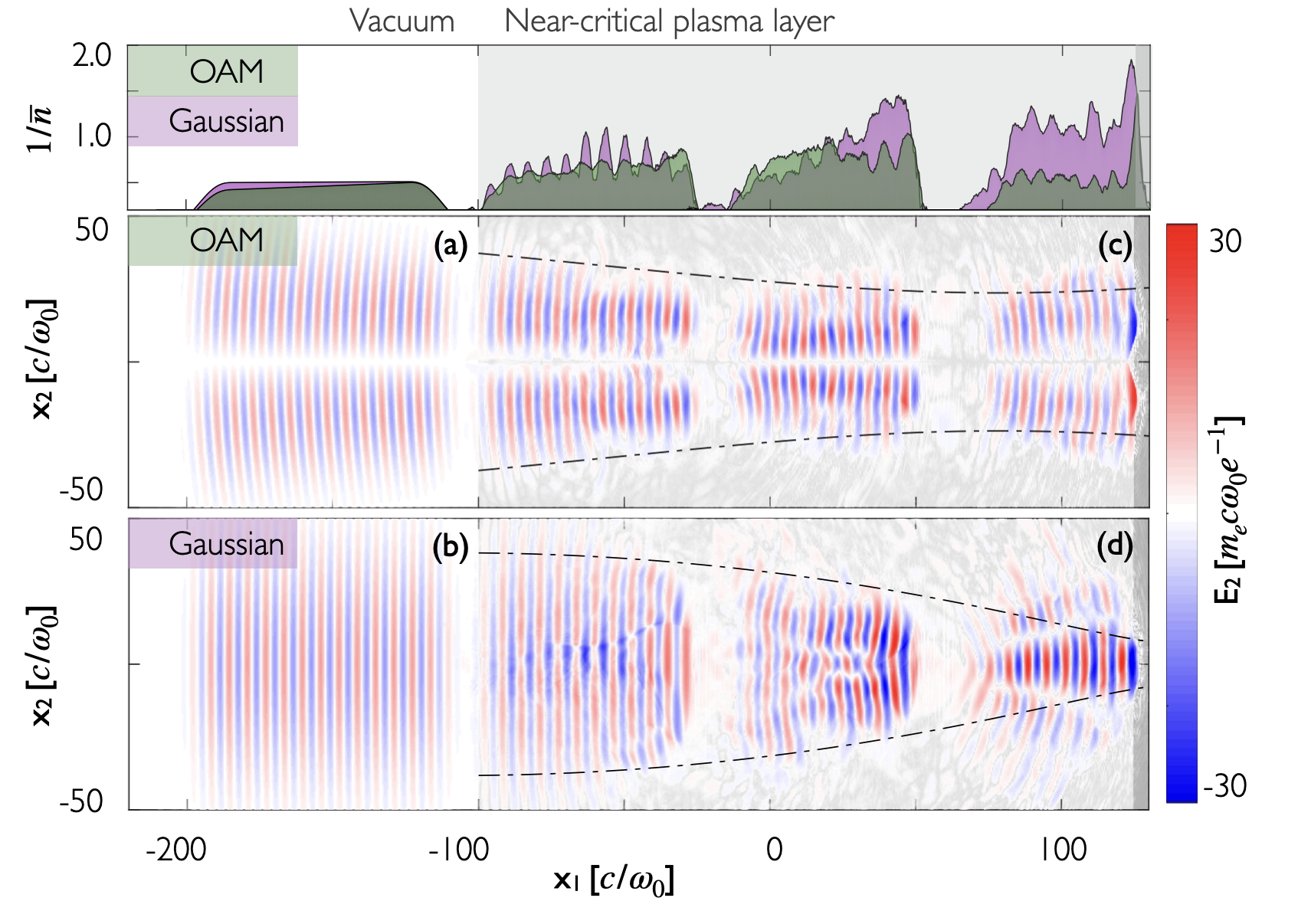}
\caption{\label{fig:selffocus}Laser pulse at four consecutive time steps at the relative propagated position, demonstrating the differences in relativistic self-focusing between Gaussian and OAM pulses. The near-critical plasma layer starts at $-100 c/\omega_0$ and ends at $125 c/\omega_0$. The laser electric field component of OAM and Gaussian lasers are in (a- Vacuum, c - Near-critical plasma) and (b - Vacuum, d - Near-critical plasma), respectively, and above the corresponding inverse of the relativistic transparency factor $1/\overline{n} = \gamma n_{cr}/n_e$ for OAM (green) and Gaussian (purple) vector potentials. The superimposed dashed line corresponds to the analytical solution of the beam waist evolution, calculated with Eqs. (\ref{eqn:dgleq}) and (\ref{eqn:enveq}).}
\end{figure}

Theory predicts ponderomotive channeling of the Gaussian beam leading to an enhanced focusing, where the laser beam width reduces by a factor of $\sim 3$ close to the end of the near-critical layer. In contrast, due to its reduced critical power ratio, the OAM beam focuses slowly, with a maximum increase in spot size by $\sim 1.5$. After that, the beam width diverges gradually.
The prediction for the different relativistic self-focusing behavior agrees with the simulation results and can be observed in Fig. \ref{fig:selffocus}.
Figure \ref{fig:selffocus} (a) and (b) show the initial laser pulse in vacuum and panels (c) and (d) show three consecutive positions of the modulated laser pulse when propagating through the near-critical plasma. The superimposed dashed line corresponds to the analytical solution from Eqs. (\ref{eqn:dgleq}) and (\ref{eqn:enveq}) and demonstrates that the predicted trend for the beam waist evolution matches the simulation results. Figure \ref{fig:selffocus} (c) demonstrates a slightly increased amplitude of the OAM driver (except for the first $\lambda/2$ of the pulse front at the last position due to photon deceleration \cite{tsung2002}) and also pulse compression. Figure \ref{fig:selffocus} (d) shows a longitudinal pulse compression of the Gaussian pulse with a strong amplitude enhancement by a factor of $\sim 3-4$ and surrounding filamentation patterns. Note that the power is not conserved due to laser energy absorption. Both Gaussian and OAM drivers can trap and accelerate background plasma electrons. However, accelerated electrons only become radially confined in the OAM driver case.
In the top panel of figure \ref{fig:selffocus}, the corresponding evolution of the inverse of the relativistic transparency factor, i.e. $1/\overline{n}$, for the thin overdense layer is presented. The laser amplitude of the Gaussian laser pulse has grown enough such that $1/\overline{n} \gtrsim 1$; thus, the laser breaks through the overdense layer. In contrast, the overall laser amplitude of the OAM driver stays efficiently small such that $1/\overline{n} < 1$. Consequently, the laser with OAM is reflected from the thin overdense layer, leaving a cold rear surface. As a direct result, the proton acceleration is mainly driven by the cylindrical symmetric sheath fields generated by the hot electrons from the near-critical plasma layer escaping into a vacuum.

\begin{figure}[h!]
\includegraphics[scale = 0.35]{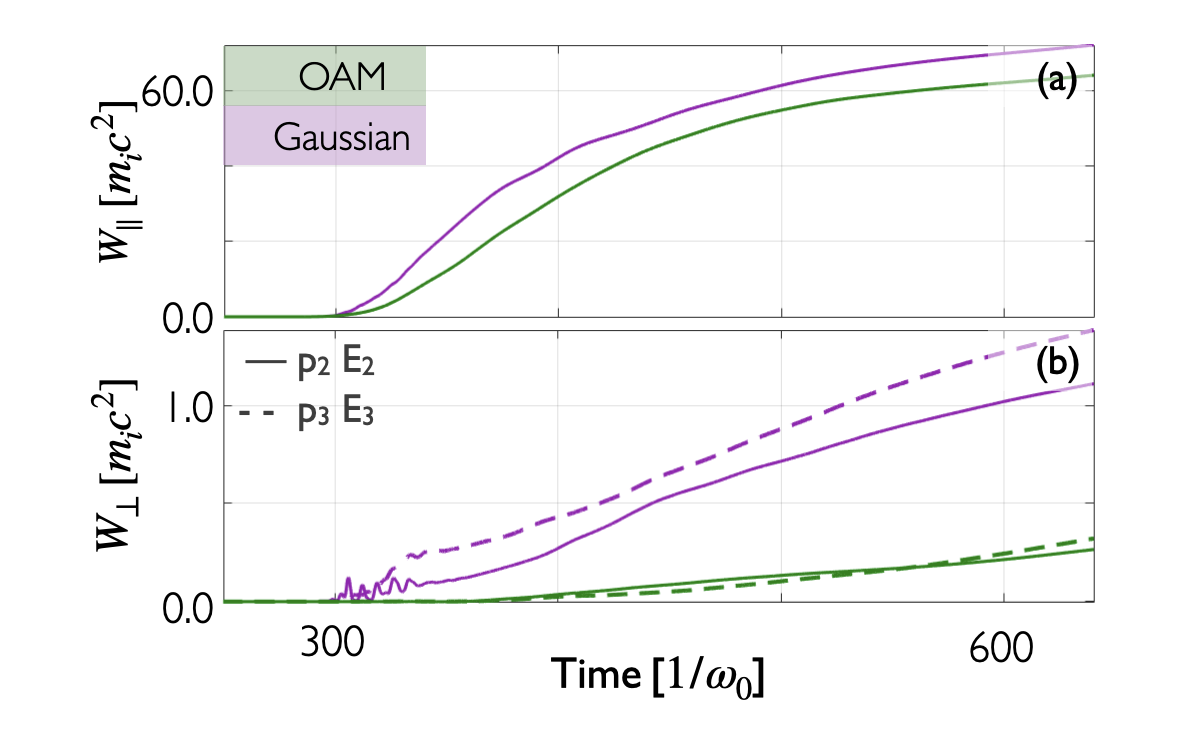}
\caption{\label{fig:work} Averaged time evolution of the cumulative sum of the work components $W = \int p_j(t)/(\gamma m_i) \cdot E_j(t) dt$ acting on selected protons after irradiation with Gaussian (purple) and OAM (green) driver with (a) the longitudinal work component $W_\parallel = p_1 \cdot E_1$, and (b) the transverse work components $W_\perp = p_\perp \cdot E_\perp$, where $p_2\cdot E_2$ corresponds to the laser polarization direction. In the OAM scenario, the transverse work acting on the protons is significantly smaller.}
\end{figure}

A strong and cylindrical symmetric longitudinal sheath field is essential to pull out protons with a small angle from the backside of the target.
The influence of each sheath electric field component on the proton acceleration can be examined through the evolution of the work acting on the protons, determined by $W =\int p_j(t)/(\gamma m_i) \cdot E_j(t) dt$, where $\gamma = \sqrt{1+(\vert \vec{p} \vert/m_i c)^2} \approx 1$, $t$ is the time and $E_j$ is the electric field component indicated by $j={1,2,3}$ for the longitudinal, laser polarization and transverse direction, respectively. 
We focus on the most energetic protons only, i.e. we select the protons with a longitudinal momentum $p_1>0.2 \, m_i c$. 
The proton energy gain due to the longitudinal field component $E_1$ is the strongest and evolves similarly for  Gaussian and OAM drivers (Fig. \ref{fig:work} (a)), which indicates, among other things, a similar hot electron generation in the near-critical plasma layer. Fig. \ref{fig:work} (b), which depicts the time evolution of the transverse work components, shows that, in the OAM case, these protons with the highest energies are significantly less influenced by the transverse electric field components. This is because the protons are pulled out centered where the longitudinal sheath field is the strongest and keep a small angle due to its cylindrical symmetry. On the contrary, in the Gaussian case, we observe periodic energy oscillations with the laser period until ($\sim 300$ $1/\omega_0$), indicating a direct influence of the laser due to the onset of RIT. In addition, during the main acceleration phase and due to the asymmetry of the sheath fields, the transverse work done on the selected protons grows much faster in the Gaussian case than the OAM case, which will inevitably increase proton bunch divergence when using a Gaussian driver. 

\begin{figure}[h!]
\includegraphics[scale = 0.33]{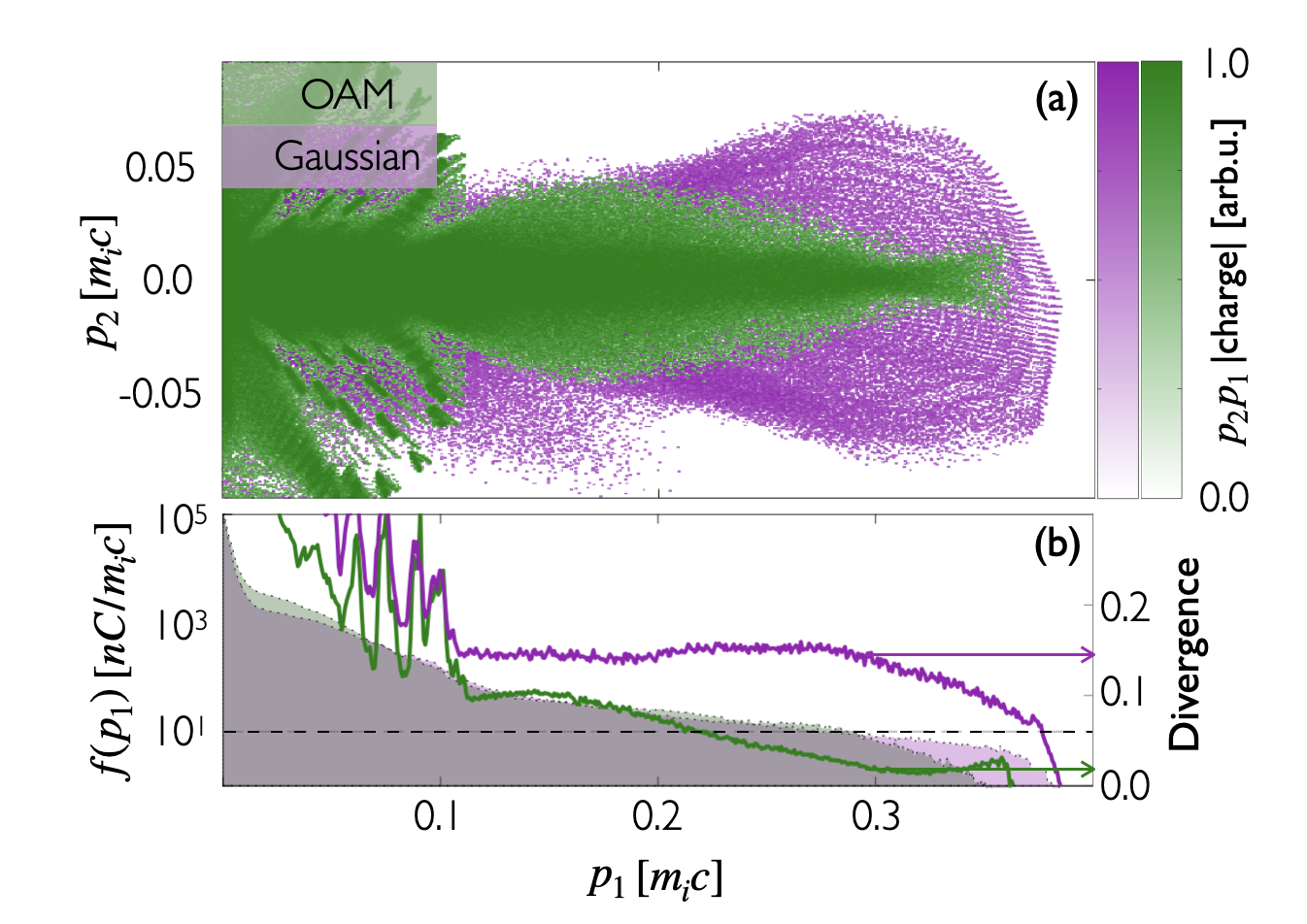}
\caption{\label{fig:intch} Properties of proton bunches $\sim 93$ fs ($520$ $1/\omega_0$) after Gaussian (purple) and OAM (green) drivers reached the overdense plasma layer, show significantly smaller divergence in the OAM case and a similar longitudinal charge distribution. (a) $p_2-p_1$ momentum space of forward accelerated protons. Dashed line marks the cutoff at $0.01 \%$ of the maximum of the integrated charge function $f(p_1)$. (b) Corresponding integrated charge over transverse momentum $f(p_1)$ (left axis - area plots) and divergence (right axis - lines). The arrows indicate the value of divergence at the cutoff ($p_1 \approx 0.3 m_i c$ ).}
\end{figure}

The work done by the transverse electric field components $E_\perp$ on the protons has a direct impact on the proton bunch divergence as shown in Fig. \ref{fig:intch} that indicates a strong reduction of the proton bunch divergence when using the OAM driver, in comparison to the Gaussian case. The image is taken at time 93 fs ($520$ $1/\omega_0$) after the laser pulses have reached the overdense plasma layer of the double-layer target.  
Figure \ref{fig:intch} (a) illustrates the momentum space $f(p_1,p_2)$ in laser polarization ($p_2$) and positive longitudinal ($p_1$) direction. We observe a collimated proton bunch with strong axial symmetry only in the scenario of an OAM driver. Figure \ref{fig:intch} (b) shows the corresponding integrated charge (area plots) on the left axis and the divergence, defined by $\text{Divergence}(p_1)= \sqrt{ \int (p_2-p_{2Centroid})^2 f(p_1,p_2) dp_2/\int f(p_1,p_2) dp_2}/p_1$, on the right axis, where $p_{2Centroid} = \int p_2 f(p_1,p_2) dp_2/\int f(p_1,p_2) dp_2$. We measure the cutoff energy at $0.01 \%$ of the maximum integrated charge $f(p_1)$, highlighted by the superimposed dashed line since the charge becomes insignificant for lower values. The conversion efficiency from laser to forward accelerated protons at the rear side is similar, with  $\approx 5\%$.
The integrated charge $f(p_1)$ evolves similarly in both cases up to a cutoff at $p_1 \lesssim 0.29 m_i c$ ($E_\mathrm{cutoff} \approx 40$  MeV). For $E_\mathrm{cutoff} \pm 10\%$, we obtain an accelerated proton charge of $\approx 0.3$ nC in both cases. However, the divergence of the accelerated proton bunches (at the cutoff energy) differs significantly by a factor of $\sim 6.5$. 
Hence, this novel setup leads to proton bunches with similar energies and charge, but while the beam spreads in the case of a Gaussian driver, it is highly collimated in the case of an OAM driver.

\section{General conditions for the production of low divergence proton bunches through self-focusing dynamics} 
\begin{figure}[h!]
\includegraphics[scale = 0.35]{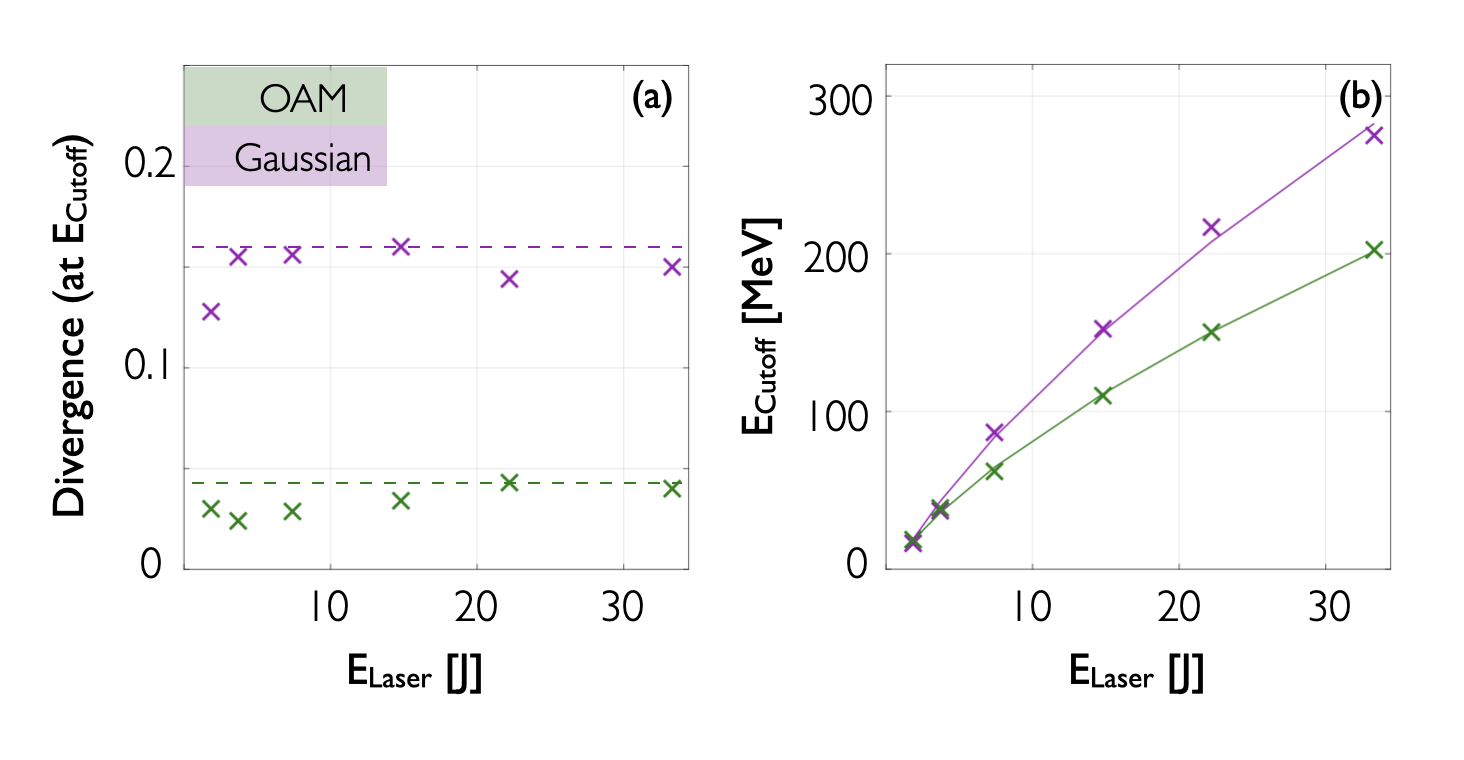}
\caption{\label{fig:laserene_comp} Dependence on the laser pulse energy of Gaussian and OAM drivers of: (a) the proton bunch divergence at the cutoff energy where the dashed line highlights the maximum values of 0.04 (OAM) and 0.16 (Gaussian), (b) the corresponding proton energy at cutoff where the cross relates to simulation results and the line to a fit function $E_\mathrm{cutoff}\propto a_0 \, \log(n_1)$.
In each simulation the plasma densities ($n_1,n_2$) of the double-layer target increase by the same factor as the the laser amplitude $a_0$. The cutoff energy $E_\mathrm{cutoff}$ is measured at $0.01 \%$ of the maximum of the integrated charge function $f(p_1)$.}
\end{figure}

Figure \ref{fig:laserene_comp} demonstrates that our findings hold over a broad range of laser energies (2-33 J), as long as the laser amplitude ($a_0$) and plasma densities $n_1$ and $n_2$ scale appropriately. In order to preserve the physics we discussed, the relativistic transparency factor $\overline{n}$ has to stay constant. 
Thus, we scaled $a_0 \rightarrow c_I \, a_0$ and $(n_1, n_2) \rightarrow c_I \, (n_1,n_2)$, where $c_I$ is the scaling factor, assuming that $\gamma \propto a_0$ for $a_0 \gg 1$. 
In Fig. \ref{fig:laserene_comp} (a), we find that for the driver with OAM the proton divergence stays low with values $\lesssim 0.04$ while the divergence is broader with values going up to $\sim 0.16$ in the Gaussian case. We measured the cutoff energy at  $0.01\%$ of max($f(p_1)$) which results in a proton charge of $\gtrsim 0.3$ nC ($>10^{10}$ protons) for $E_\mathrm{cutoff} \pm 10\%$. In Fig. \ref{fig:laserene_comp} (b), we demonstrate that the cutoff proton energy $E_\mathrm{cutoff}$ increases gradually for laser energies ranging from $\sim 2-33$ J depending on both the laser amplitude $a_0$ and the electron density $n_1$. We find that the approximated relation $E_\mathrm{cutoff} \approx T_h \, \left[\log(n_{h0}/\tilde{n})-1\right]$, presented in Ref. \cite{pazzaglia2020}, is in good agreement with our simulation results, assuming a ponderomotive-like electron heating $T_e \propto a_0$ (concordant to ponderomotive and superponderomotive electrons,  discussed in Refs. \cite{cialfi2016,DLT_Bin}) a linear scaling of the hot electron density with the near-critical plasma density $n_{h0} \propto n_1$ and a constant free parameter $\tilde{n}$ (used in ref. \cite{pazzaglia2020}).
 
\section{Conclusions}
In conclusion, the proposed concept, explored with three-dimensional PIC simulations and supported by analytical considerations of relativistic self-focusing in near-critical plasma, reveals a new approach for generating very collimated proton beams in the tens of MeV energy with moderated laser energy. The approach is based on a double-layer target configuration to exploit the benefits of improved laser energy absorption and on a laser with OAM to exploit its reduced relativistic self-focusing and its cylindrical symmetry. 
The results have been obtained with a set of realistic laser parameters that indicate the same longitudinal energy ($\sim 40$ MeV) and charge ($\sim 0.3$ nC) for both the Gaussian and OAM laser drivers, with a strong reduction of beam divergence, by a factor of $\sim 6.5$, in the OAM case. We have demonstrated through various simulations that our findings hold over a broad range of laser energies ($2-33$ J) by relying on an appropriate scaling between laser intensity and target densities, with a proton cutoff energy that scales as $E_\mathrm{cutoff} \propto a_0 \, \log(n_1)$, and protons with a peak energy of more than 200 MeV that can be delivered with a 30 J (PW) laser system. We believe that even when adjusting the target configuration or the energy of the Gaussian driver, we cannot achieve the same reduction in divergence while maintaining the energy gain of the proton bunch as the OAM case has demonstrated. 
This study opens new opportunities for conceiving compact proton sources for applications requiring a collimated and energetic proton beam.

\begin{acknowledgements}
The project that gave rise to these results received the support of a fellowship from ”la Caixa” Foundation (ID 100010434). The fellowship code is “LCF/BQ/DI19/11730025”. We acknowledge PRACE for access to resources on MareNostrum (Barcelona Supercomputing Center). The work was supported by the European Research Council (InPairs ERC-2015-AdG no. 695088), by The Schwartz/Reisman Center for Intense Laser Physics, by a research grant from the Benoziyo Endowment Fund for the Advancement of Science, by the Israel Science Foundation, Minerva, Wolfson Foundation, the Schilling Foundation, R. Lapon, Dita \& Yehuda Bronicki, and by the Helmholtz association.
\end{acknowledgements}

\appendix
\section{APPENDIX: Study of the spot size evolution under varying initial conditions} \label{appendix:relfoc}
\begin{figure}[h!]
\includegraphics[scale = 0.3]{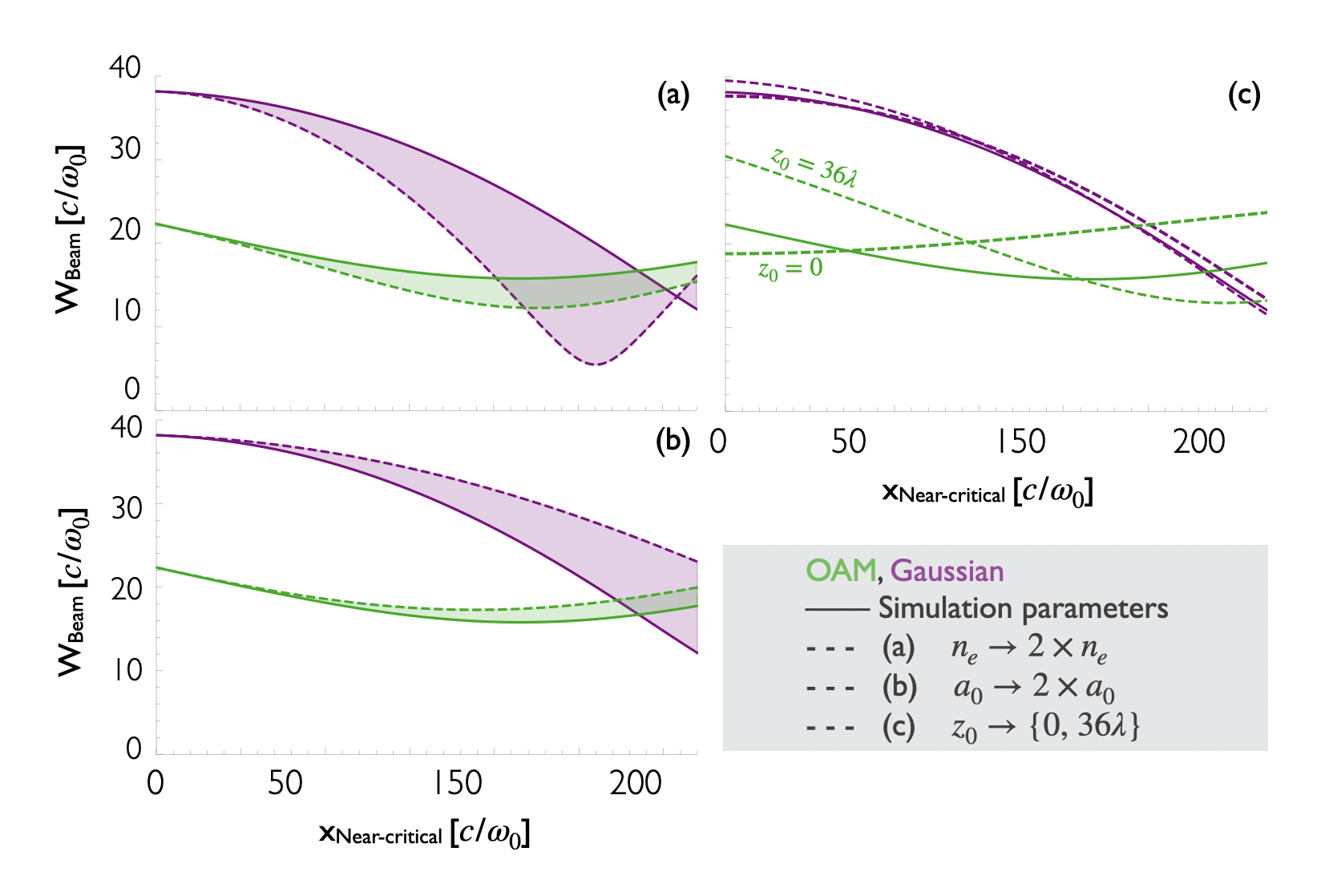}
\caption{\label{fig:suppmat1} Relativistic self-focusing of laser with OAM is less sensitive to adjustments in electron density $n_e$ or laser amplitude $a_0$ but responds more to shifts of the focus than Gaussian laser. Three different scenarios are considered, in a) the electron density $n_e$ is doubled, in b) the laser amplitude $a_0$ is doubled and in c) the position of the focus is moved to the beginning of the near-critical plasma layer, i.e. $z_0 =0 $, and to the end of the plasma layer, i.e. $z_0 \approx 220 c/\omega_0$, with reference to the parameters from the initial simulation case.}
\end{figure}

We find less sensitivity of the twisted laser mode's beam width evolution on the plasma density $n_e$ and the laser amplitude $a_0$ but a dependency on the position of the focal spot $z_0$ in comparison to a Gaussian mode.
We compare the beam width evolution of a pure Gaussian mode and a twisted laser mode with radial index $p=0$ and azimuthal index $\ell = 1$ by solving the envelope equations for the laser spot size Eqs.(\ref{eqn:dgleq})-(\ref{eqn:enveq}). We take into account that the critical power ratio of the twisted laser mode is reduced by a factor of four, i.e. $P/P_c = \frac{1}{4} P/P_G$, and the initial conditions of our simulation setup. Note that the initial spot sizes of the Gaussian mode and the twisted laser mode differ already by a factor of two to keep the laser energy and the weighted averaged amplitude equal. 
In figure \ref{fig:suppmat1}, we show that the twisted laser mode is less sensitive to changes in the plasma density $n_e$ (Fig. \ref{fig:suppmat1} (a)) and laser amplitude $a_0$ (Fig. \ref{fig:suppmat1} (b)), and more sensitive on the initial position of the focal spot $z_0$ (Fig. \ref{fig:suppmat1} (c)) which is related to the initial focal spot size. In Fig. \ref{fig:suppmat1} (c), we show that the beam width evolution of the twisted laser changes least for a focal spot position in the center of the near-critical layer $z_0 \approx 110 c/\omega_0$ which we find to be most favorable for the presented setup. The relativistic self-focusing of the Gaussian laser is barely affected by the initial focal spot position. 


\newpage
%
\bibliography{bibtexversion}

\end{document}